\documentclass[11pt]{article}
\usepackage{graphicx} %
\usepackage[margin=1in]{geometry}
\usepackage{url}
\usepackage{hyperref}
\usepackage[colorinlistoftodos]{todonotes}

\newenvironment{summary}{%
    \clearpage
    \section*{Summary}
    \addcontentsline{toc}{section}{Summary}
}{%
    \par\bigskip
}

\title{The HEAL Data Platform}
\author{Brienna M. Larrick$^1$, PhD, \and
	 	L. Philip Schumm$^1$, MA, \and
		Mingfei Shao$^1$, MS,\and
		Craig Barnes$^1$, PhD, \and
		Anthony Juehne$^{1,2}$, MPH, \and
		Hara Prasad Juvvla$^1$, MS, \and
		Michael B. Kranz$^3$, PhD, \and
		Michael Lukowski$^1$, MS, \and
		Clint Malson$^1$, BS, \and
		Jessica N. Mazerik$^{2,4}$, PhD, \and
		Christopher G. Meyer$^1$, PhD, \and
		Jawad Qureshi$^1$,BS \and
		Erin Spaniol$^{2,5}$, MPH, \and
		Andrea Tentner$^3$, PhD, \and
		Alexander VanTol$^1$, BS, \and
		Peter Vassilatos$^1$, MS, \and
		Sara Volk de Garcia$^1$, PhD, \and
		Robert L. Grossman$^{1,6, *}$, PhD \protect \\
       \protect \\
    $^1$Center for Translational Data Science, University of Chicago, Chicago, Illinois, USA\\%
    $^2$This work was completed while this coauthor was at NIH\\%
    $^3$NORC, University of Chicago, Chicago, Illinois, USA\\%
    $^4$Independent Contractor, USA\\%
    $^5$US Department of Veterans Affairs, Washington, DC, USA\\%
    $^6$Dept. of Medicine \& Computer Science, Univ. of Chicago, Chicago,  Illinois, USA \\%
}

\date{
   September, 2026
}

\newcommand{\ph}[1]{\medbreak \noindent {\bf #1}}

\begin{document}

\maketitle

\bigbreak
\ph{*Corresponding Author:} Robert L. Grossman, Center for Translational Data Science, University of Chicago, 900 East 57th Street, Chicago IL 60637, USA; robert.grossman@uchicago.edu; 773-702-5660

\bigbreak 
\ph{Keywords:} data mesh, data ecosystem, FAIR data sharing, secondary data analysis, secure research environments

\begin{summary}
The different projects in the NIH Helping to End Addiction Long-term (HEAL) Initiative\textregistered\ have generated a wide variety of data, including clinical trial data, EHR data, patient-reported outcomes data, behavioral and psychological assessments, preclinical and animal studies data, omics data, imaging data, and public health surveillance data. These datasets have ended up in over two dozen different NIH and non-NIH data repositories. The HEAL Data Platform is a data mesh enabling users to search and discover data across the repositories and enabling authorized users to securely analyze them. With a data mesh, data are not centralized, but instead remain in their respective data repositories until they are accessed by an authorized user or working group for analysis. This work is relevant to efforts by NIH and other scientific agencies to increase data sharing and maximize the value of existing repository infrastructure through use of data meshes.

\end{summary}

\clearpage

\section*{Abstract}

\ph{Objective.}
The objective was to develop a cloud-based, federated system to serve as a single point of search, discovery and analysis for data generated under the NIH Helping to End Addiction Long-term\textregistered \ (HEAL) Initiative. 

\ph{Materials and Methods.}
The HEAL Data Platform is built on the open source Gen3 platform, utilizing a small set of framework services and exposed APIs to interoperate with both NIH and non-NIH repositories. Framework services include those for authentication and authorization, creating persistent identifiers for data objects, and adding and updating metadata.

\ph{Results.}
The HEAL Data Platform serves as a single point of discovery of over one thousand studies funded under the HEAL Initiative. With hundreds of users per month, the HEAL Data Platform provides rich metadata and currently interoperates with nineteen data repositories and commons to provide access to shared datasets. Secure, cloud-based compute environments that are integrated with STRIDES facilitate secondary analysis of HEAL data. 

\ph{Discussion.}
Studies funded under the HEAL Initiative generate a wide variety of data types, which are deposited across multiple NIH and third-party data repositories. The mesh architecture of the HEAL Data Platform provides a single point of discovery of these data resources, accelerating and facilitating secondary use.  

\ph{Conclusion.}
The HEAL Data Platform enables search, discovery, and analysis of data that are deposited in connected data repositories and commons. By ensuring that these data are fully Findable, Accessible, Interoperable and Reusable (FAIR), the HEAL Data Platform maximizes the value of data generated under the HEAL Initiative.

\clearpage

\section{Introduction}

The HEAL Data Platform (https://healdata.org) is a cloud-based, federated data ecosystem that supports and accelerates research on pain and opioid use disorder in support of the NIH HEAL Initiative \cite{collins2018helping}.  As of August 2026, the Platform includes over 1,100 HEAL-supported studies, interoperates with 19 different data repositories and data commons, and provides the ability to analyze completed and released studies in secure workspaces. In this paper, we describe the design, implementation, and operations of the system.

\ph{NIH Data Sharing Policy and the HEAL Initiative.}
It is well recognized that the sharing of high-value datasets can accelerate new scientific discoveries and  enable the validation of research results \cite{nelson2009data}. An important step forward in requiring data sharing was the 2023 NIH Data Management and Sharing Policy \cite{jorgenson2021incentizizing}. The policy requires a data management and sharing plan for NIH-funded research and promotes the accessibility and reuse of data. For most research, this means depositing data in one of the data repositories or data commons operated by NIH Institutes and Centers, or one of the third-party data repositories that are used by the scientific community.

For many projects, programs and initiatives, the data and derived data fit naturally into a single data repository.  On the other hand, the NIH HEAL Initiative \cite{collins2018helping} supports over 1,000 projects and programs whose aim is to improve the understanding, management and treatment of pain, as well as the management and prevention of opioid misuse and addiction, and produces a range of data and derived data that are ultimately deposited in multiple NIH and third-party data repositories and data commons.  The goal of the HEAL Data Platform is to enable researchers to search for and discover studies and datasets of interest from the NIH HEAL Initiative, and to use cloud-based workspaces (analysis environments) to analyze the data or to download or transfer the data for analysis.

\ph{Data repositories, commons, hubs, and ecosystems.}
In this paper, it is useful to distinguish between data repositories, data commons, data hubs, and data ecosystems.  \emph{Data repositories} archive and distribute data deposited by investigators and vary between those providing no data curation services and those performing robust data curation. A \emph{data commons} also archives and distributes data but includes additional services designed to ensure that the data are FAIR (e.g., indexing, metadata, and data governance services), to permit users to analyze data (e.g., cloud-based workspaces), and to permit interoperability with other systems (e.g., open application programming interfaces (APIs)); these are often targeted to a specific research community \cite{grossman2022ten-lessons}.  A \emph{data hub} interoperates with several data repositories and commons to provide a single web-based portal where users can search for and access data from a range of disparate sources, often delivering additional services such as enhanced metadata and search capabilities and cloud-based workspaces. Importantly, in general, a hub does not store data, but only provides pass-through access to users who have already been authorized according to the data source's own procedures. Finally, a \emph{data ecosystem} or \emph{data mesh} consists of multiple data repositories and/or data commons, possibly including a data hub, together with cloud-based metadata stores, knowledge bases, and search and computational resources that can interoperate \cite{grossman2024safe, grossman2024annotated}. 

\ph{The HEAL Data Ecosystem and HEAL Data Platform.}
The HEAL Data Ecosystem is a data ecosystem containing multiple data repositories and data commons, and including a data hub (provided by the HEAL Data Platform) that connects the ecosystem together and provides workspaces enabling authorized researchers to access and analyze data. 
While hubs can be organized in a variety of ways, the HEAL Data Platform hub is organized around the concept of a study, which can be associated with one or more datasets and one or more publications.  Alternatively, data hubs may be organized around datasets, as is the case for the Biomedical Research Hub (e.g. \cite{barnes2021biomedical}); around patients that are gathered from multiple different studies; around publications and their associated data; or around a particular data type, such as DICOM files (e.g. the MIDRC Biomedical Imaging Hub). 

Important functions of a hub in a data ecosystem include the following:

\begin{enumerate}
    \item Providing a central portal that researchers can use to discover and access data of interest across two or more data repositories or commons; and
    \item Providing a place to attach workspaces (also known as authorized analysis environments) that can be used to securely analyze data that a researcher discovers via the hub.
\end{enumerate}

\noindent Critical to enabling these functions is that the participating data commons and repositories contain APIs that expose metadata and data so that the datasets they host can be discovered and accessed. The data hub aggregates the \emph{metadata} to be used as the basis of search and discovery, while the \emph{data} themselves remain in their respective commons or repositories, with access governed by the commons or repositories own, existing procedures.

The HEAL Data Platform is one of the first data meshes for use by NIH-funded studies, and it required the development and implementation of several unique features.  Unlike a hub that automatically aggregates information about {\em all} available data from specific repositories, the HEAL Data Platform was intended to facilitate access to and secondary analysis of data from studies related to a specific initiative: the NIH HEAL Initiative. This is a disparate group spanning several NIH institutes, scientific disciplines, research phases, and study designs, and therefore their data are being deposited into at least two dozen very different repositories. To accommodate this, we started with a list of HEAL-funded awards, added study-specific metadata from NIH RePORTER and the HEAL Program Office, and loaded these into the Platform's Metadata Services. We then added a form permitting an investigator to ``register'' their study with the hub by claiming ownership of its metadata record, linking it to other existing sources of metadata (e.g., ClinicalTrials.gov), and indicating in which repository the data was or will be be deposited. While it is standard for investigators to provide some of this information to a repository at the time they are depositing their data, soliciting their help when constructing a hub and permitting them to determine how their studies appear on the hub has, to our knowledge, not been done before.

The repositories into which HEAL data are being deposited vary considerably in their metadata and data models (including generalist repositories that don't have data models), their procedures for data submission and governance, and whether they have an API that permits interoperability as part of a data mesh. As a result, we had to be extremely flexible, at least initially, in how we approached interoperability. In cases where a repository's API was limited or non-standard, we implemented a temporary procedure (when possible) to provide data access. For repositories without a data API, we asked them to create a cloud bucket to which we could connect and, in the case of controlled access data, a corresponding allow list. As most repositories that did not have a fully capable API were in various stages of developing one, participating in the HEAL Data Ecosystem provided both an opportunity for them to reassess the design and capabilities of their future API and for us to provide technical suggestions and advice.

\section{Methods}

\subsection{HEAL Data Platform Architecture}

\ph{Gen3 mesh services.}  The HEAL Data Platform was built on the open source Gen3 data platform (\url{https://Gen3.org}), which relies on a small number of software microservices (called {\em Gen3 Data Mesh Services or Framework Services}) that allow the HEAL Data Platform to interoperate and form a mesh with data repositories, data commons, and other data and computational resources with open APIs.  Designed to enable data to be fully FAIR, the Gen3 mesh services include open APIs and services for authenticating and authorizing users; controlling user access to data objects via a policy engine; creating persistent identifiers for data objects; and adding, accessing and updating metadata. The Gen3 user interface facilitates execution of search and discovery of studies by users, and is configurable to expose underlying metadata. This architecture is an example of the end-to-end design principle that is the basis of the internet architecture \cite{saltzer1984end}, which is sometimes called the narrow middle architecture \cite{grossman2018progress, grossman2025proposed} in the context of data ecosystems.  

\ph{Gen3 Metadata Services.}
Gen3 provides two metadata APIs for use by data commons and data ecosystems that are used to manage the metadata for the HEAL Data Platform. The Gen3 Metadata APIs provide a flexible and highly scalable key-value store that can accommodate arbitrary, semi-structured metadata, as well as an API for submitting and querying those metadata. They also permit the harvesting of metadata from existing/external sources, which can then be harmonized and used to support queries across multiple data repositories. New sources are added by creating an adapter to harvest the metadata (ideally via an API provided by the source, though scraping is also possible) and perform whatever harmonization is required. Once the adapter is in place, ingesting and updating metadata from the source is performed automatically according to a specified schedule. Metadata from Gen3 Metadata Services may be used to support search or other user interfaces, and may be accessed programmatically via API.

\begin{figure}
    \centering
    \includegraphics[scale=0.60]{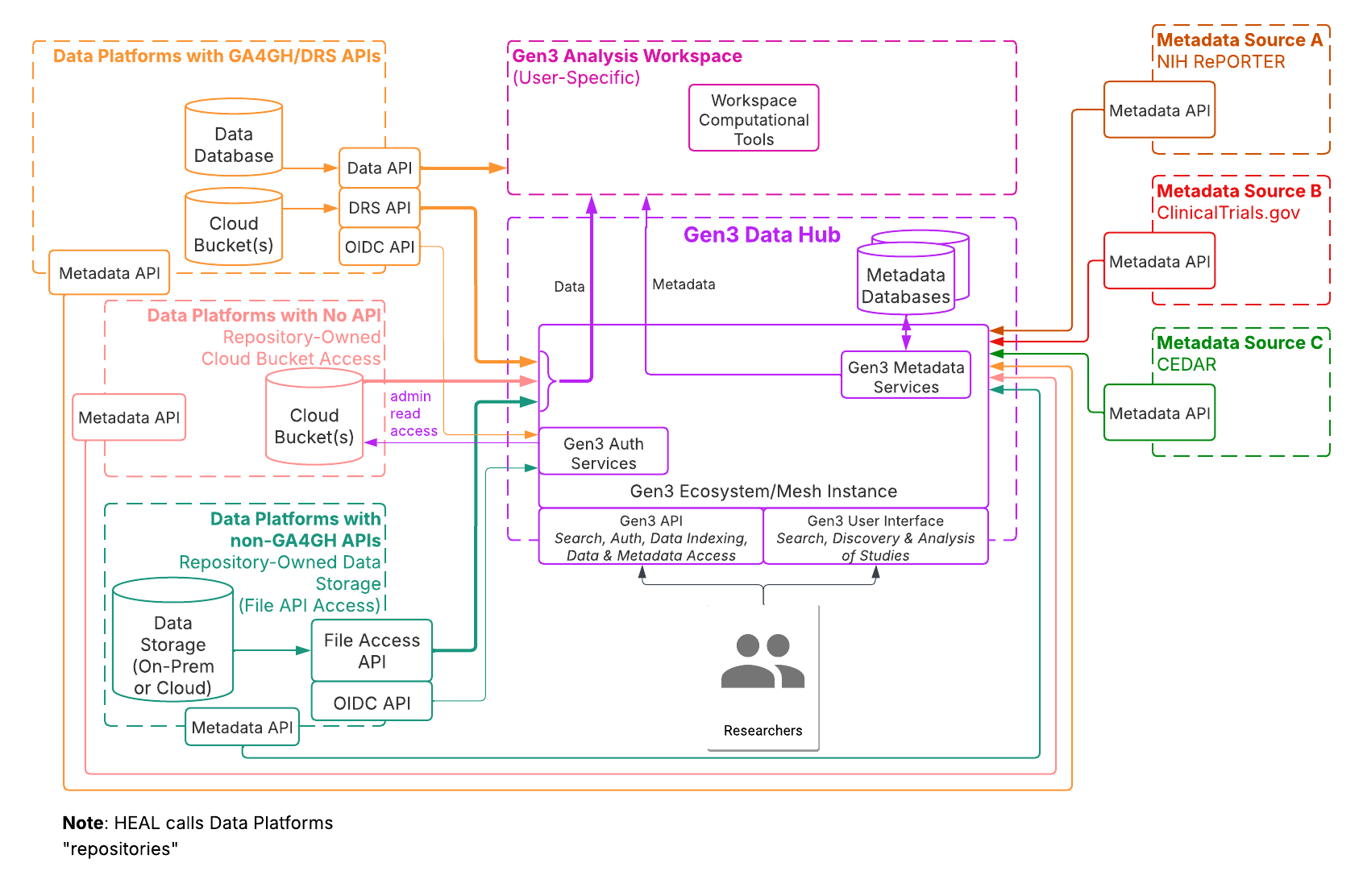}
    \caption{The HEAL Data Platform architecture.}
    \label{fig:heal-architecture}
\end{figure}

With these services, the HEAL Data Platform collects metadata from a variety of sources, such as NIH RePORTER, ClinicalTrials.gov, data repositories and individual investigators via the Center for Expanded Data Annotation and Retrieval (CEDAR) Workbench (https://metadatacenter.org) \cite{martinez2017supporting}. See Figure~\ref{fig:heal-architecture}.

\subsection{HEAL Data Platform Security and Compliance Framework}

We use the term HEAL Data Platform {\em Core} to refer to the HEAL Data Platform Hub, mesh services, and workspaces and the term {\em connected data repositories} to refer to the connected data repositories, data commons, and other connected third party systems that are part of the HEAL Data Platform. With this terminology, the HEAL Data Platform consists of the HEAL Data Platform Core and the connected data repositories.  

The Core and connected data repositories interoperate following the {\em platform governance} principles described in \cite{grossman2024safe}, which were the result of the NIH Cloud Platform Interoperability Program (NCPI) Community/Governance Working Group discussions that took place during 2021-2023.  Broadly speaking, platform governance can be thought of as complementing data governance principles and providing a basis for two cloud platforms operated by two different organizations to interoperate. 

The HEAL Data Platform Core operates at a Federal Risk and Authorization Management Program (FedRAMP) Moderate level, following the relevant FedRAMP requirements \cite{fedramp2025docs} and the Moderate policies, procedures, controls and other requirements described in NIST SP 800-53 \cite{nist800-53}. 

Each connected data platform is responsible for the data governance, platform governance, security, privacy, and compliance of the hosted data resource, including agreements for data contributors and data use, and decisions affecting security, compliance, and interoperability.

The security and compliance between the Core and the connected data repositories is described in interoperability agreements between the University of Chicago that operates the Core and the organizations that operate the various connected data repositories following NIST SP 800-47 \cite{dempsey2021managing}.  The HEAL Data Platform calls these ``System Interoperability Agreements (SIA).'' These agreements follow NIST SP 800-47 \cite{dempsey2021managing}.  The SIAs describe how the Core and repositories interoperate with each other following the platform governance principles described in \cite{grossman2024safe}.

\subsection{Interaction with Data Repositories}

\subsubsection{General Principles}

The HEAL Data Platform interoperates with generalist and domain-specific data repositories to serve as a single point of discovery and access of data generated under the NIH HEAL Initiative. Broadly speaking, for a data platform to join a data mesh: i) the governance committee for the data mesh must approve the platform; ii) the data platform must implement and support the minimum technical requirements for interoperability; iii) and, as mentioned above, the data platform and the data mesh sponsor must each sign an SIA that specifies how they interoperate. As described in \cite{grossman2024pillarsdatameshes}, there is a set of minimal requirements a data repository or commons must meet to join a mesh:
\begin{enumerate}
    \item Data in a data platform should be identified with Persistent Identifiers (PIDs);
    \item Data in a data platform should have FAIR APIs so that the metadata associated with a PID can be accessed;
    \item Data in a data platform should have FAIR APIs so that the data associated with a PID can be accessed;
    \item There should be an API for authentication and authorization.
    \item  A data platform in a data mesh should provide an API so that authorized analysis environments can access data on behalf of authenticated users who are authorized to access data in the platform;
\end{enumerate}

Providing the hub in a data mesh, the HEAL Data Platform relies on a hybrid governance model, where each connected data platform has its own organizational governance structure, with shared governance in some areas such as minimum metadata. As such, in accordance with the principles laid out in \cite{grossman2024pillarsdatameshes}, the HEAL Data Platform:
\begin{enumerate}
    \setcounter{enumi}{5}
    \item Agrees with connected repositories on a shared governance model covering data governance and platform governance;
    \item Agrees with connected repositories on the types of data objects supported, and the minimum metadata required for each type of data object;
    \item Assigns a PID to each object supported;
    \item Returns usage statistics and usage information of who used the data hub to access data to the data repository that provides the data;
    \item Provides a public API interface to the HEAL Metadata Services for the data mesh metadata.
\end{enumerate}

\subsubsection{Variation Across the Repository Landscape}

While the section above describes the general principles by which the HEAL Data Platform interoperates with data repositories, the details vary depending on a repository's API capabilities, its administrative policies, and how it stores and distributes its data. When we began work on the HEAL Data Platform in 2020 few repositories had data access APIs, though several had the development of such APIs on their roadmap. Our challenge was to accommodate as many repositories as possible at the outset, while anticipating future development on their part and allowing for smooth transitions as they added new API capabilities. Since then, the HEAL Data Platform has supported and collaborated with several repositories as they have worked to add new APIs.

Broadly speaking, we encountered four types of repositories. The first group (represented in the orange box in Figure~\ref{fig:heal-architecture}) had or were in the process of adding APIs for data access, including APIs for authentication and authorization (AuthN/AuthZ), based on the Global Alliance for Genomics and Health (GA4GH) API specifications; specifically, their Data Repository Service (DRS) and Passport specifications \cite{rehm2021ga4gh}. These specifications are compatible with NIH's Researcher Auth Service (RAS) Initiative, and are supported by the Gen3 platform out of the box. Unfortunately, repositories that do not archive genetics and genomics data are currently unlikely to maintain such APIs.

A second group of repositories have AuthN/AuthZ and data access APIs that do not follow the GA4GH specifications but adhere to other standards (e.g., OAuth 2.0/OIDC). Among HEAL repositories that maintain data access APIs this is the largest group, and is represented in the green box in Figure~\ref{fig:heal-architecture}). Interoperating with these repositories typically required some amount of custom development on our part, as well as continuing development to maintain interoperability if and when the repository changes its API in order to maintain interoperability. An important subset of this group are those that distribute open, unrestricted data via open URLs, since in this case we were able to index the files using Gen3's indexing service and then make them accessible without any custom work. Data requiring only that the user create an account with the repository (so-called ``registration-only'' access) can also sometimes be handled within this group if the files are accessible by the HEAL Data Platform via URLs (or a simple API) and the Platform's authentication service and access logs (information from which may be shared with the repository) meet the repository's requirements for limiting and monitoring data access.

Repositories that do not have data access APIs or that have non-standard and/or limited APIs constitute a third group.  In some cases, these repositories are in the process of upgrading their capabilities which would permit them to move to one of the first two groups.  Thus, we designed an interoperability strategy that would be immediately accessible to any repository without imposing any technical or resource requirements, and that could serve as a ``bridge'' for those repositories in the process of implementing standard AuthN/AuthZ and data access APIs.  Briefly, the repository creates a dedicated cloud bucket of which they are the owner and into which they put copies of data files generated by HEAL-funded studies (as shown in the pink box in Figure~\ref{fig:heal-architecture}). The HEAL Data Platform is then granted read access to the bucket so that it may index the files and permit approved users to access them.  Finally, authorization information is transferred from the repository to the Platform which then stores this information and handles the authentication and authorization using Gen3's built-in services.  We continue to work individually with each repository in this group to identify and maintain a secure and efficient way to transfer the necessary authorization information.  As with the second group, this strategy can also be used to implement ``registration-only'' access.

A fourth and final group of repositories are those that do not store and distribute data in the form of persistent, indexable files, or whose administrative policies preclude any form of direct interoperability. An example of the former are repositories that manage their data in one or more databases, and allow approved users to query those databases in order to build a user-specific, custom data package containing only those data required for a specific analytic purpose.  For repositories in this group, the HEAL Data Platform negotiated an agreement that allows users to retrieve data from the repository (using its existing procedure(s)) and upload them to a workspace within the Platform for analysis.  Efforts to identity a more efficient and secure method for transferring data from the repository to a Platform workspace are ongoing.

\subsection{Direct Interaction with Data Creators/Investigators}
One of the key features enabling search and discovery of studies and data through the HEAL Data Platform is a rich set of study- and variable-level metadata. The HEAL Data Platform accomplishes this in part by utilizing APIs to aggregate metadata from public sources (e.g., NIH RePORTER and the National Library of Medicine's ClinicalTrials.gov). However, the key element in enabling this is interacting directly with data providers (study investigators) to curate these metadata before data have been submitted to repositories. The Platform features a ``registration'' service that enables authenticated users to request and be granted permissions to submit and modify study- and variable-level metadata corresponding to a specific study record on the Platform. These services allow for multiple users to collaborate in the metadata submission process, and for study leads to delegate tasks to members of their teams.

\section{Results}

\subsection{The HEAL Data Platform}
\begin{figure}
    \centering
    \includegraphics[scale=0.30]{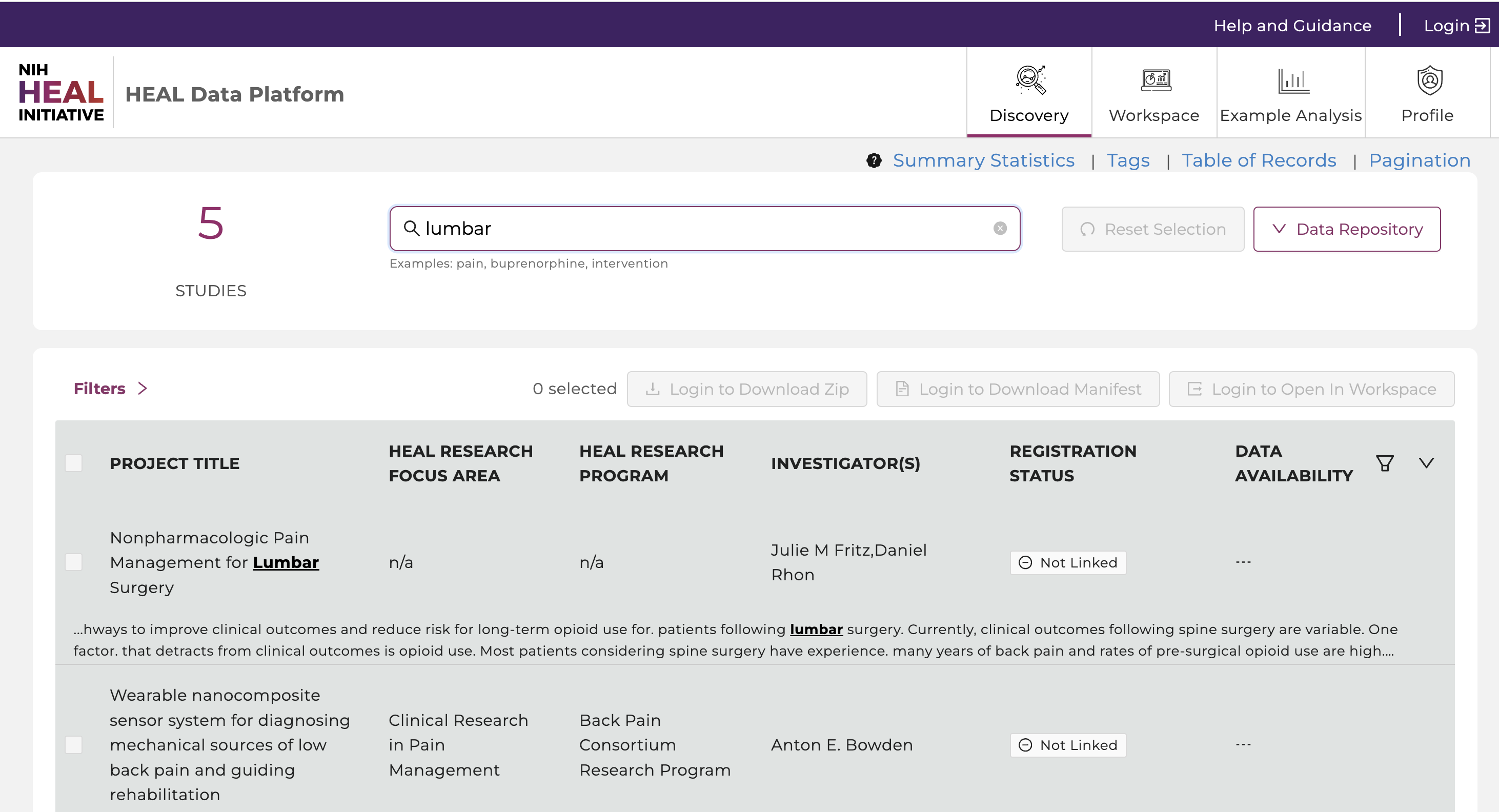}
    \caption{The HEAL Data Portal}
    \label{fig:portal}
\end{figure}

The HEAL Data Portal (see Figure~\ref{fig:portal}) is a Gen3 Data Hub and Analysis Workspace that enables users to search for datasets across the over two dozen data repositories that have been ingested into a Gen3 Metadata Service.  

\begin{figure}
    \centering
    \includegraphics[scale=0.30]{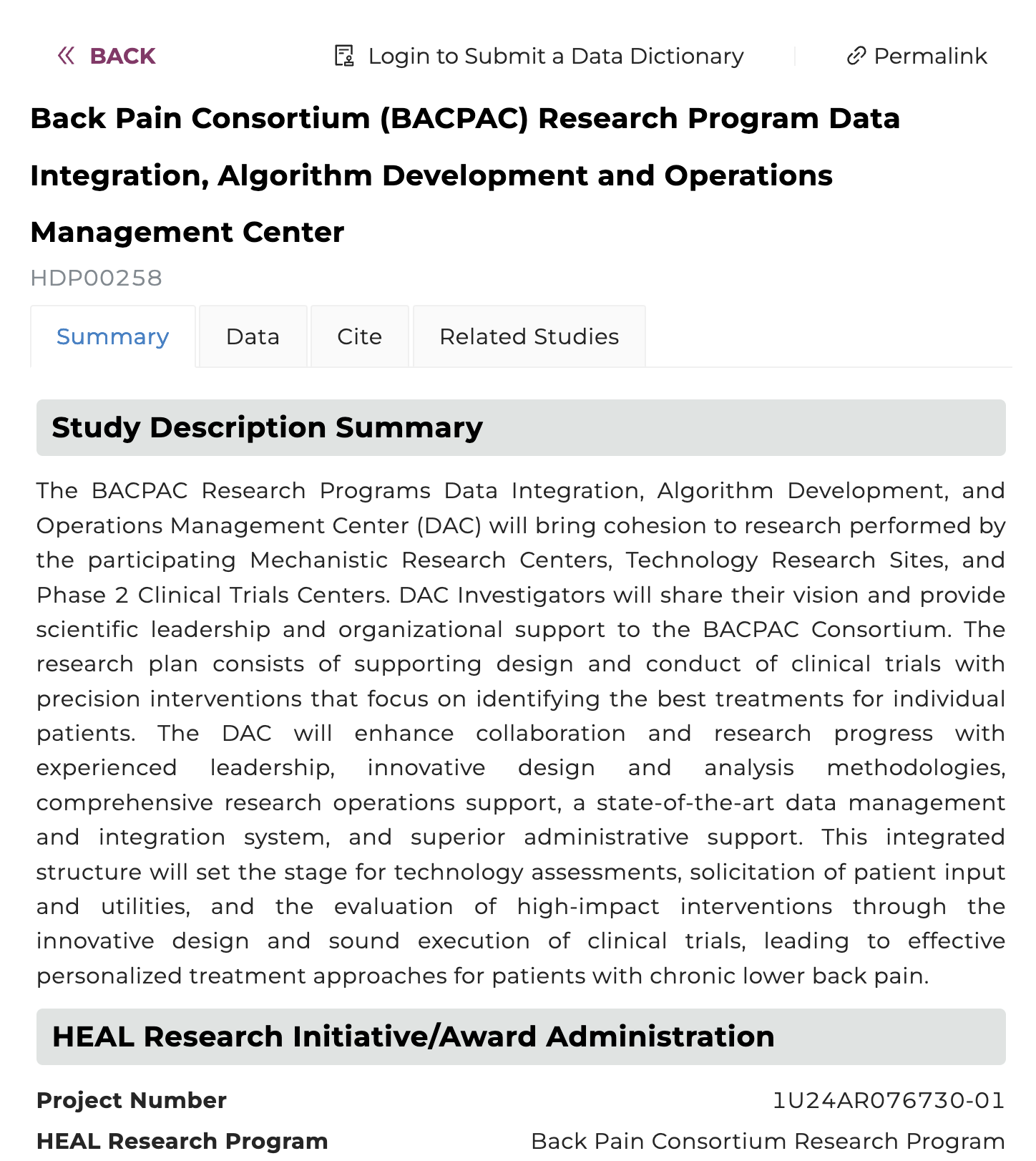}
    \caption{Additional information about a study can be seen once the study is registered by the PI.}
    \label{fig:study-detail}
\end{figure}

As mentioned above, at the time of publication, the HEAL Data Platform contains over 1,100 studies. See Table~\ref{tab:overview}.  Information about studies is added in two key steps. In the first step, information is gathered automatically from the NIH RePORTER \cite{nih2024reporter} through its API. Next, projects and programs funded by HEAL are asked to register their study with the HEAL Data Platform. This registration step enables two key things that support the collection of a robust set of study-level metadata. First, the provision of an NCT number by the investigator during registration enables the Platform to query the \url{ClinicalTrials.gov} API, maintained by the NIH National Library of Medicine, and ingest the available metadata into the HEAL Metadata Services. Second, the registration process results in the instantiation of a study-level metadata (SLMD) form, into which the investigator can provide information about the study in a standardized way. The HEAL Data Platform leverages the CEDAR platform for this purpose. The PI can then also add a data dictionary, enabling search across the elements in the data dictionary, including over any HEAL Common Data Elements \cite{sheehan2016improving}.  Figure~\ref{fig:study-detail}.

\subsection{HEAL Data Platform  Workspaces}
The HEAL Data Platform offers secure, cloud-based workspaces so users can compute over data that are discovered and accessed through the Platform. The workspaces use Gen3 mesh services for user authentication and authorization, and for retrieving data objects that are accessed from connected data repositories. Users can import data from one or more connected repositories for secondary analysis, and/or upload data files from their local machine. HEAL workspaces include Jupyter notebooks and support analysis in Python, R and Stata, and come pre-installed with commonly-used packages, plugins and utilities. 

HEAL workspaces are integrated with the NIH STRIDES Initiative (https://cloud.nih.gov), and users can request access to a workspace and apply for STRIDES credits. While the request for STRIDES credits is processing, the user is granted immediate "trial" access to get started using HEAL workspaces. 

\begin{table}[t]
    \centering
    \begin{tabular}{|l|r|} \hline
        Searchable studies & 1,198 \\ \hline
        Connected data repositories & 19 \\ \hline
        Registered studies & 616 \\ \hline
        Studies with study-level metadata & 511 \\ \hline
        Studies with variable-level metadata & 99 \\ \hline        
        Available datasets & 161 \\  \hline
    \end{tabular}
    \caption{An overview of the HEAL Data Platform as of August, 2026}
    \label{tab:overview}
\end{table}

\subsection{HEAL Study-Level Metadata} 

Gen3's Metadata Services can store and query a wide range of semi-structured (i.e., key-value) metadata. Typically, those metadata are obtained from existing sources or generated by data curators. The HEAL Data Platform obtains metadata from sources such as NIH RePORTER and \url{ClinicalTrials.gov}, as well as from repository APIs when available. At the same time, the requirement noted above to provide a rich search and discovery experience \textit{before data have been submitted to repositories} required us also to solicit metadata directly from investigators. To support this, we augmented Gen3's existing Metadata Services to permit user submission of metadata in two ways: (1) by uploading metadata files through the Platform, and (2) by utilizing the CEDAR Workbench and API to permit form-based entry of metadata.

Study-level metadata — information describing the study such as its objectives, research design, participant population, experimental and data collection methods, etc. — permit users to browse and search across studies, and richer metadata contribute to a more efficient and effective search experience. Because most existing metadata standards are discipline-specific, none of these were adequate to accommodate the full range of HEAL-funded studies, and a decision was made at the program level to develop a new, general metadata model for use by the HEAL Data Platform. This new model was informed by existing standards (e.g., the Dublin Core) and designed to accommodate the full range of HEAL studies while maintaining consistency whenever possible with the data repositories into which the data are being deposited.

Although some \emph{administrative metadata} are generated at the time a study is funded, the majority of \emph{descriptive metadata} are often generated at the time a study is nearing completion in anticipation of reporting the results and submitting the data to a repository. Due to the urgent nature of the opioid crisis and the fact that most HEAL-funded studies were still in progress at the time the HEAL Data Platform was deployed, we started soliciting study-level metadata from investigators immediately in order to permit users of the Platform to search over studies. These metadata may be subsequently updated, such as when a study's data are being submitted to a repository.

\subsection{HEAL Variable-Level Metadata}
Along with a model for study-level metadata, the HEAL Ecosystem also developed a schema for variable-level metadata to facilitate data harmonization across studies. The schema is built on the Frictionless framework, and is purposefully lightweight so that it can be applicable to the diverse array of data generated under the HEAL Initiative. Both the HEAL variable- and study-level metadata schemas can be found on GitHub (https://github.com/HEAL/heal-metadata-schemas).

All metadata on the Platform are completely open, with the idea that when data are being packaged for submission to a data repository, the metadata that have been accumulated by the HEAL Data Platform can be submitted as part of that package. Similarly, these metadata are available for secondary users or repositories themselves to view and access directly.

\subsection{Hybrid Governance}    

Each data repository in the HEAL mesh is operated by a separate organization and has its own governance. In addition, the HEAL Data Platform (specifically, the HEAL hub) has its governance.  We use the standard term {\em hybrid governance} to refer to the shared data governance and platform governance that describe how the data repositories in the data mesh and the data hub interoperate. One of the challenges of the project was developing agreements that enabled the multiple data repositories to participate in the HEAL ecosystem. The standard agreement covers how the two systems will interoperate with API calls, what reporting the hub should provide to a data repository when data are accessed from the data repository via the hub, what happens in the case of a data breach, and other issues.  We eventually developed an agreement that covered most common use cases, which greatly reduced the time required to negotiate agreements between the hub and a new data repository joining the mesh.

\section{Discussion}

Maximizing the scientific, clinical, and policy value of a large and interdisciplinary research initiative such as HEAL requires effective sharing of the data and other materials generated by its funded studies, ideally in a way that facilitates searching across all studies funded by the initiative as well as using the resulting data, especially pooling data across studies. One possible strategy for accomplishing this would have been to build a dedicated repository (or data commons) for archiving HEAL data. However, this strategy would have been wasteful in not using the many existing data repositories, including those maintained by NIH and others. %
Many of the HEAL-compliant repositories chosen to meet high standards for data quality and stewardship are domain-specific with specialized data models, curation procedures, and governance protocols. These not only facilitate submission of certain types of data but also serve to reassure investigators that their data are being handled and distributed appropriately. Even investigators submitting their data to generalist repositories that do not provide common data models or curation services can benefit from using a repository with which they are familiar and/or for which their institution provides support. By interoperating with a broad range of repositories to provide centralized access to data and metadata, the HEAL Data Platform makes efficient use of existing data sharing resources, adds capabilities to accelerate and facilitate the secondary use of HEAL data, and provides secure analysis environments that repositories without their own workspaces may refer their users to.

We encountered some technical challenges with this approach to data access. Many repositories did not have APIs and of those that did, most either did not use standard, up-to-date protocols (e.g., OIDC/OAuth2.0 for authentication and authorization) or were incomplete (e.g., provided a metadata API but no data API). This prevented us from immediately interoperating with these repositories in a straightforward and standards-compliant manner. Instead, whenever possible we designed a targeted, temporary strategy to provide data access until the repository is able to develop or expand its API. In our experience such repositories were appreciative of the information and guidance we provided based on our experience with other repositories, and were often willing to plan their development to include the FAIR APIs necessary to interoperate with platforms like the HEAL Data Platform.

\section{Conclusion}

The HEAL Data Platform is a data ecosystem comprised of a data hub and analysis workspace that connects to data repositories and commons operated by NIH Institutes and Centers, and third-party data repositories that are used by the scientific community. Through the use of APIs that expose metadata and data, and a small number of framework services for creating persistent identifiers for data objects, user authentication and authorization, and adding and updating metadata, the HEAL Data Platform enables search and discovery of data in connected data repositories. While repositories and commons maintain responsibility and control over the governance of their hosted data resources, the HEAL Data Platform enables authorized users to access datasets, and to analyze them in cloud-based workspaces which come pre-installed with commonly-used packages, plugins and utilities, and support analysis in Python, R and Stata.  

By facilitating FAIR data sharing and secondary use of data, data ecosystems maximize the scientific and analytic value of research data. With over 1,100 studies across over two dozen data repositories, the HEAL Data Platform supports the NIH-wide effort to speed scientific solutions to stem the national opioid public health crisis.

\section{Resource Availability}
\subsection*{Lead contact}
Requests for further information and resources should be directed to and will be fulfilled by the corresponding author, Robert Grossman (robert.grossman@uchicago.edu).

\subsection*{Materials availability}
\begin{itemize}
    \item Study- and variable-level metadata schemas were developed and are available at \url{https://github.com/HEAL/heal-metadata-schemas}
    \item Variable-level metadata files for HEAL Common Data Elements (CDEs) are available in the HEAL Metadata Service at \url{https://healdata.org/mds/metadata?data=True&limit=1000&_guid_type=cde_metadata}
    \item Software tools for generating and using variable-level metadata are available at \url{https://github.com/uc-cdis/heal-platform-sdk}
\end{itemize}

\subsection*{Data and code availability}
\begin{itemize}
    \item The HEAL Data Platform is available at: \url{https://healdata.org/} Access to data and metadata is available through FAIR APIs.
    \item The HEAL Data Platform is built with the Gen3 software framework (\url{https://gen3.org/}) for constructing data commons and meshes, which is open source and available at \url{https://github.com/uc-cdis}
\end{itemize}

\section{Acknowledgements}
This work is supported by the National Institutes of Health (NIH) under award number 3 OT2 OD030208-01S4. We thank the data repositories that participate in the mesh and the HEAL investigators who have shared metadata and archived their data in the connected repositories.

\section{Contributions}

BML, LPS, RLG contributed to the conception and high level design of the HEAL Data Platform. BML, LPS, MS, CB, AJ, HPJ, MBK, ML, CM, JNM, CGM, JQ, ES, AT, AV, PV, SVG, RLG contributed to the development and operations of the HEAL Data Platform.

\bibliographystyle{vancouver}
\bibliography{refs-23.bib,refs-24.bib}

\end{document}